# New solution of the compressible Navier-Stokes equation

## Sergey G. Chefranov[1),2)] and Artem S. Chefranov[1),3)]

[1)] Obukhov Institute of Atmospheric Physics, Russian Academy of Sciences, Moscow, Russia
[2)] Physics Department, Technion-Israel Institute of Technology, Haifa 32000, Israel
[1)] schefranov@mail.ru , [2)] csergei@technion.ac.il , [3)] a.chef@bk.ru

## Abstract

We use the general exact solution of the Cauchy problem for the compressible Euler vortex equation in unbounded space which was obtained earlier (S.G.Chefranov, Sov. Phys. Dokl., **36**, 286, 1991). This solution loses its smoothness in finite time and coincides with the exact solution of the Hopf equation, describing the inertial motion of the ideal fluid without pressure. On this base we obtain here the new smooth at all times solution to the compressible Navier-Stokes (NS) equation with the pressure field shows linear proportionality to the divergence of the velocity field, as it is known for an out-of-equilibrium systems with large second viscosity and small first viscosity. For example, directly from this solution of the NS equation for the case of two-dimensional compressible flow the exact representation of energy spectrum $E(k) \propto O(1/k^3)$ is obtained.

## Introduction

In laboratory experiments, as well as in numerical experiments, revealed is a significant non-Gaussianity of turbulence, manifested as intermittency in the region of small scales for the distribution of the energy viscous dissipation rate value [1-5]. In this regard, considered is the mechanism of intermittency due to the implementation of singular behavior in the evolution of enstrophy, which determines the value of the turbulence energy dissipation rate [1, 5]. However, because of the nonlinearity of the hydrodynamic equations, it has not been possible so far to obtain a closed description of the evolution in time of enstrophy and higher moments of the vortex field, which would allow estimating the degree of deviation of the turbulent flow statistics from the Gaussian one [2] and determining the existence of a finite limit for the value of the turbulence energy dissipation rate at a limiting small viscosity.

In this paper we have obtained a response to these two interrelated questions on the base of a new class of exact solutions for the equations of hydrodynamics of an ideal compressible medium which was obtained earlier [6]. It is established here that near the singularity of the solution the difference of the asymmetry and excess values from the ones typical to the Gaussian process is clearly evident. This follows from the obtained exact expressions for the vortex field one-point moments. More over the energy spectra for two-point correlation functions are also obtained on the base of the mentioned above exact solution [6], which is modified by introduction of effective viscosity and homogeneous friction. Here, instead of the traditional use of the medium equation of state, from the integral entropy balance equation, a linear relationship between the pressure field and the divergence of the velocity field of a viscous compressible medium is derived. It is typical to an out-of-equilibrium system with large second viscosity [7, 8] (see (81.4), (81.6) in [7] and (1.95) in [8]). Obtained is the necessary and sufficient condition of appearance of singularity in the vortex three-dimensional (3D) Euler equation solution in a finite time and shown a possibility of regularization of this solution at the expense of any small viscosity or super-threshold homogeneous friction. The homogeneous friction is always arising in any numerical simulations of the Navier-Stokes equation when truncation procedure is used at some large wave numbers or some small size of discretization for network of space resolution. The obtained here new smooth at all times solution of the compressible Navier-Stokes equation (see also [9-12]) corresponding to a positive solution of the Millennium prize problem generalization for the case of compressible flows, for which only a negative solution of the problem of the existence of a smooth solution on an arbitrarily large time interval has a priori been assumed earlier for the velocity field with finite integral kinetic energy at initial time t=0 and for t>0 [13] (www.claymath.org ).

1. **Euler and Hopf equations and their exact vortex solution**

1. The Euler equations, which express the impulse and mass conservation laws, for the case of ideal compressible medium are well-known for already more than 250 years (since 1755) and have the following now day form [14, 15]:

$$\frac{\partial u_i}{\partial t} + u_j \frac{\partial u_i}{\partial x_j} = \frac{1}{\rho}(f_i - \frac{\partial p}{\partial x_i}) \quad , \quad (1.1)$$

$$\frac{\partial \rho}{\partial t} + \frac{\partial}{\partial x_i}(\rho u_i) = 0 \quad , \quad (1.2)$$

where according to the repeated indices meant is a summation from 1 to n (n is a dimension of space), and $u_i; f_i; p; \rho$ the components of the velocity field, the external forces field, field of pressure and density, respectively. If we consider not only external forces, but also the force associated with the medium viscosity, then equation (1.1) coincides in form with the Navier-Stokes equation, known since 1827 [7].
Indeed, equation (1.1) exactly coincides with the Navier-Stokes equation for viscous compressible medium, if for the force $f_i$ in (1.1) the following representation takes place [7, 16]:

$$f_i = \eta \frac{\partial^2 u_i}{\partial x_k^2} + (\varsigma + \frac{\eta}{3}) \frac{\partial}{\partial x_i}(\frac{\partial u_k}{\partial x_k}) \quad (1.3)$$

In (1.3) $\eta; \varsigma$ are the constant in space coefficients of the first viscosity and the second viscosity, respectively.

2. Since 2000 the problem of the existence of smooth solutions for the three-dimensional (3D) Navier-Stokes equation is one of the seven fundamental problems for the Millennium prize formulated at the Clay Mathematical Institute [13]. In [13] this problem is formulated not for the general form of the Navier-Stokes equation (1.1), (1.3), but for the special case when the approximation of an incompressible medium $\rho = const$ is assumed to be fulfilled in (1.1) - (1.3) for zero divergence of the velocity field

$$div \vec{u} \equiv \frac{\partial u_k}{\partial x_k} = 0 \quad , \quad (1.4)$$

A necessary condition for carrying out this approximation is the assumption that the Mach numbers $Ma = \frac{|\vec{u}|}{c} \ll 1$ are small (where c is the speed of sound in a given medium).

Such a formulation is connected not only with the explicit simplification of the form of system (1.1) - (1.3). The main thing here is the a priori idea that the complete system (1.1) - (1.3) cannot have smooth solutions on an arbitrarily large time interval. The reason for this is the possibility of realization of a singularity (collapse) arising in the solution in a finite time $t_0 > 0$, as, for example, in case of collapse of a traveling nonlinear wave in an ideal compressible medium [17, 18]. At the same time it is assumed that the consideration of the viscosity forces cannot lead to the regularization of the corresponding vortex solutions. To understand whether this is really so, it is necessary to have at least some class of exact non steady vortex solutions of the Euler equations for the case of compressible medium.

3. Indeed, L. Euler noted the complexity of the analysis necessary to obtain the general form of the solution of system (1.1), (1.2) and pointed out the importance of obtaining at least particular solutions of these equations [14]. Thus, for example, in [14] (see also [15]), considered is the solution corresponding to an exact hydrostatic equilibrium, when in (1.1) the total force on the right side of equation (1.1) is equal to zero and all the velocity field components are identically equal to zero. In [14] also considered is the case when at the same zero balance of forces on the right side of (1.1) the velocity is no longer zero, but equal to a constant value that is the same for all Lagrangian fluid particles in the medium.

Following to this logic of Euler's work [14], and also under the condition of zero balance of forces on the right side of the Euler equation (1.1), we can consider that despite the velocity is constant for each Lagrangian fluid particle, it is not necessarily strictly the same for different particles of the medium, which may have different initial velocity. Besides, the requirement of equality to zero of the left side of equation (1.1) leads to the following nonlinear equation for the nonzero and not constant (as in [14]) velocity field in the case of zero balance force in the right side of (1.1):

$$\frac{\partial u_i}{\partial t} + u_k \frac{\partial u_i}{\partial x_k} = 0, i = 1,...,n \quad (1.5)$$

This equation describes inertial motion of the medium particles and is the n- dimensional generalization of the well-known one-dimensional equation for nonlinear travelling wave, sometimes called as the Burgers equation at zero viscosity [17], as the Hopf equation [18], or as the Riemann equation [6]. Indeed, in one-dimensional case equation (1.5) is obtained within the limit of large Mach numbers $Ma \gg 1$ from equation, firstly considered by Riemann [7] (see equation (101.2) in [7]).

The closed system of equations of type (1.5), (1.2) is also obtained from the equations of hydrodynamics of self-gravitating dust matter when considering the formation of a large-scale structure of the Universe under the assumption of zero pressure $p = 0$ in (1.1) [17, 19, 20]. The same system of equations (1.5), (1.2) is also obtained when expanding the Mach numbers $Ma \gg 1$ in inverse powers, for example, in the hydrodynamic description of the granular media dynamics, when the contribution to (1.1) associated with the pressure gradient is a small value of order $O(1/Ma^2), Ma \gg 1$ [21- 23].

4. A special interest attracts just the vortex solution of the equation (1.5), for which a possibility for realization of the enstrophy explosive growth is caused by not only the vortex filament 3D stretch effect, but the vortex field generation at the expense of finiteness of the velocity field divergence in the compressible medium flow.

Actually, after applying the curl operator to the equation (1.5) the Euler vortex 3D equation (also called the Helmholtz vortex equation) follows from it and has the following form:

$$\frac{\partial \omega_i}{\partial t} + u_k \frac{\partial \omega_i}{\partial x_k} = \omega_k \frac{\partial u_i}{\partial x_k} - \omega_i \frac{\partial u_k}{\partial x_k}, \quad (1.6)$$

where $\vec{\omega} = rot\vec{u}$. Exactly the same equation is obtained when applying the curl operation to initial Euler equation (1.1) (when in (1.1) $rot\vec{f} = 0$ and fulfilled is the condition of parallelism between the pressure and density field gradients $rot(\frac{1}{\rho}\vec{\nabla}p) = 0$). Equation (1.6) is sometimes called the Friedmann equation just for the case of a compressible medium.

Thus, a class of solutions to the equation (1.5), having a nonzero vortex field, definitely determines a class of equation (1.6) solutions for the vortex field of the compressible ideal medium flow.

However, all well-known solutions of the Hopf 3D equation (1.5) are obtained only in Lagrange variables [17-19]. Therefore they do not give the solution for the Euler vortex 3D equation (1.6) for the vortex field just in Euler variables. At the same time, for both the problems of astrophysics, considered in [17,19, 20] and hydrodynamics of granular media noted is the importance of just the vortex solutions for the Hopf equation (1.5) [21, 23]. In connection with the Millennium problem consideration [13] it is also interesting to obtain the solution for just the equation for the vortex field (1.6). When it is open problem of subsequent evaluation of viscosity force effect on a possibility of regularization of the solution singularity.

Only in [6] for the first time the general exact vortex solution of equation (1.5) was obtained in Euler variables. The vortex field for which is an exact strong solution of the nonlinear non stationary equation (1.6) for arbitrary smooth initial velocity fields defined in an unlimited space (see also [9], where a more detailed derivation of this solution is given) with finite kinetic energy. On the basis of this solution, in the following sections we obtain an analytic smooth solution of the 3D Navier-Stokes equations (1.1), (1.3) for any times. For this solution the presentation for pressure is obtained from the entropy balance equation and in the limit of small viscosity $\eta \to 0, \eta << \varsigma$ (when second viscosity is large as for far out-of-equilibrium systems [7]) is the same as in formula (81.4), (81.6) in [7] and formula (1.95) in [8] for the deviation of pressure from it equilibrium value).
\

## 2. Energy balance and entropy equation

1. The system of Euler equations for an ideal compressible medium (1.1), (1.2) and the corresponding system of Navier-Stokes equations under condition (1.1) - (1.3) are not closed, since, for example, in the three-dimensional case it consists of four equations, but, at the same time, contains five unknown functions, namely three components of the velocity field, the pressure field and the field of the medium density.

To close this system in the limit of small Mach numbers the approximation of an incompressible medium, when the density is assumed to be a constant value and equation (1.4) takes place, is used.

In the considered herein case of compressible medium usually some equation of state [7], which connects the functions of density and pressure (here and below, let us consider an isothermal medium for simplicity's sake) is used to close the system of equations. Besides, it is assumed that the medium is in a locally thermodynamic equilibrium state, when the time of relaxation to this state is much smaller than the typical time of the flow dynamics of the medium. This assumption, however, must certainly be violated in case of velocities of the medium corresponding to not small and especially large Mach numbers, and also in case of slow relaxation processes, for example, in the presence of chemical reactions [7].

Therefore, instead of introducing any equation of state, it seems more natural to obtain an additional equation for the pressure field closing the system of equations (1.1) - (1.3), as proposed in [7] for the case of slow relaxation processes (see (81.4) in [7] and (1.95) in [8]]). In the present paper we obtain an equation for the pressure field from the consideration of the integral entropy balance equation and the sufficient condition of positive definiteness for the rate of the integral entropy growth, which gives in the limit $\eta << \xi$ exactly the same equation for pressure as (1.95) in [8] and (81.4) in [7].

At first, let us obtain from (1.1)-(1.3) the following equation of the integral kinetic energy balance for the compressible medium $E = \frac{1}{2}\int d^n x \rho u^2$ flow:

$$\frac{dE}{dt} = -\eta \int d^n x (\frac{\partial u_i}{\partial x_k})^2 + \int d^n x \left[ p - (\zeta + \frac{\eta}{3}) div \vec{u} \right] div \vec{u} \quad (2.1)$$

Expression (2.1) in the case of incompressible flow coincides exactly with given in [7] (see (16.3) in [7]) and so only the equation (2.1) may serves as its generalization for the flow of a compressible medium. Indeed, it is obtained directly from equations (1.1) - (1.3) without involving any additional thermodynamic relations. At zero viscosity $\varsigma = 0, \eta = 0$ equation (2.1) is the same as balance equation for the integral kinetic energy of compressible flow which is used in [24] (see formula (2) on page 22 in [24] when the potential of external force is zero in (2)). This distinguishes (2.1) from the formula (79.1) in [7] which does not coincide with formula (2) on page 22 of [24] in the case of zero viscosity. However, in [7] formula (79.1) is called as the generalization of (16.3) in [7] for the case of a compressible medium, but this obviously is not so because (79.1) is not the same as equation (2.1) (and also not as equation (2) in [24] for the limit of zero viscosity).

For the balance equation of total energy [7] $E_h = \int d^3 x (\rho \frac{\vec{u}^2}{2} + \rho \varepsilon)$, where $\varepsilon$ is a density of the internal energy, (see the details of derivation in [9]) we have:

$$\frac{\partial}{\partial t}(\rho \frac{\vec{u}^2}{2} + \rho \varepsilon) = -\frac{\partial}{\partial x_k} \left[ u_k (\rho(\frac{\vec{u}^2}{2} + \Phi_0) + p - (\zeta + \frac{\eta}{3}) div \vec{u}) - \eta \frac{\partial}{\partial x_k}(\frac{\vec{u}^2}{2}) \right] + T(\frac{\partial}{\partial t}(\rho s) - \frac{Q}{T}), \quad (2.2)$$

$$Q = \eta (\frac{\partial u_i}{\partial x_k})^2 - \left[ p - (\zeta + \frac{\eta}{3}) div \vec{u} \right] div \vec{u}$$

As in [7], from (2.2) and condition of the total energy conservation the balance equation for density (per unit mass) of entropy $s$ is obtained:

$$\frac{\partial}{\partial t}(\rho s) = \frac{Q}{T} \quad (2.3)$$

In (23) T is the temperature.

From (2.3) we obtain the following balance equation for the integral entropy $S = \int d^3 x \rho s$:

$$\frac{d}{dt} S = \eta \int d^3 x \frac{1}{T}(\frac{\partial u_i}{\partial x_k})^2 - \int d^3 x \frac{1}{T} div \vec{u} \left[ p - (\zeta + \frac{\eta}{3}) div \vec{u} \right] \quad (2.4)$$

Comparing the form of equations (2.4) and (2.1), for the considered isothermal case (for the case $T = T_0 = const$), we obtain the following exact relation:

$$T_0 \frac{dS}{dt} = -\frac{dE}{dt} \quad (2.5)$$

In [7] (see paragraph 79 in [7]) relation (2.5) is derived as a certain fundamental relation which must be executed for any mechanical system tending to thermodynamic equilibrium from a given out-of-equilibrium state.

However, in [7] this relation is not derived directly for the hydrodynamics of a viscous compressible medium, as opposed to the present paper, where relations (2.4) and (2.1) are derived independently and lead exactly to equality (2.5). Moreover, equation (2.4) differs from the integral entropy balance equation given in [7] (see (49.6) in [7]) obtained, as opposed to derivation (2.2)-(2.4), using additional thermodynamic relations for the pressure field, which enters into the equation of hydrodynamics (1.1).

Hence the conclusion follows that the use of thermodynamic equilibrium representations for the pressure field entering the Navier-Stokes equation for a viscous compressible medium, as it has been done in [7], is problematic because of the resulting contradiction with the fundamental relation (2.5).

Indeed, the obtained in [7] expression $\frac{dS}{dt}$ (see (49.6) in [7]) does not satisfy equation (2.5) taking into account formula (2.1), which is obtained only on the basis of equations (1.1) - (1.3) without involving any thermodynamic relations.

It means that the use of the equation of state, which is usually applied [7] for closing the system (1.1) - (1.3), also gives rise to doubts when describing the flow of a viscous compressible medium, and instead an additional equation for the pressure field must be obtained.

This additional to system (1.1)-(1.3) equation is obtained here from a sufficient condition for the positive definiteness of the integral entropy (2.4) growth velocity, which has the following form:

$$p = (\varsigma + \frac{\eta}{3})div\vec{u} \quad (2.6)$$

The system of equations (2.6) and (1.1)-(1.3) is already closed. In [7, 8] (see (81.4), (81.6) in [7] and (1.95) in [8]), a similar expression for pressure in the form of a linear function of the divergence of the velocity field is also used and obtained, albeit from other considerations, with a coefficient of proportionality equal to the coefficient of the second viscosity $\varsigma$ (under condition $\varsigma >> \eta$).

Moreover, in [7] it is also proposed to consider the linear relation (81.4) between the pressure and the divergence of the velocity field instead of the medium equation of state.

When the equation (2.6) is satisfied for the pressure field, the Navier-Stokes equation (1.1), (1.3) exactly reduces to an equation of the three-dimensional Burgers' equation type, for which in the next section proposed is an approximate solution based on the assumption of the possibility of modeling the viscosity force $\frac{\eta}{\rho}\Delta\vec{u}$ in the right side of (1.1) (when (2.6) is take place) in the limit of small first viscosity $\eta \to 0; \eta << \varsigma$ either due to a homogeneous friction $-\mu u_i$, or by means of a random Gaussian velocity field $\vec{V}(t)$ of the white noise type.

### 3. Analytical solution of the Navier-Stokes equation

1. Let us find the solution of closed system (1.1) - (1.3) and (2.6) on the basis of the exact solution of the equation (1.5), (1.6), obtained in [6] and having the following form (see validation of this in Appendix A in [9]) [9-12]:

$$u_i(\vec{x},t) = \int d^n \xi u_{0i}(\vec{\xi})\delta(\vec{\xi} - \vec{x} + t\vec{u}_0(\vec{\xi}))\det \hat{A}, \quad (3.1)$$

where $\det \hat{A}$ is a determinant of matrix $\hat{A} \equiv A_{nm} = \delta_{nm} + t \frac{\partial u_{0n}}{\partial \xi_m}$, and $u_{0i}(\vec{x})$ is an arbitrary initial smooth velocity field with finite integral of energy $E_0 = \int d^n x u_0^2(\vec{x}) < \infty$. This solution preserves smoothness only for finite time $t < t_0$, for which condition $\det \hat{A} > 0$ is executed, where $t_0$ is a minimal (in space coordinates) time for which determinant $\det \hat{A} = 0$ vanishes (see (3.7) in [9]).

In particular, for one-dimensional case n=1 we have $\det \hat{A} = 1 + t \frac{du_{01}}{d\xi_1}$ and solution (3.1) exactly coincides with the solution of equation (1.5) obtained in [24, 25]. At the same time

$$t_0 = \frac{1}{\max |du_0/dx|} > 0$$

and, for example, for initial distribution $u_0(x) = a \exp(-x^2/2L_0^2), a > 0$ we have $t_0 = \frac{L\sqrt{e}}{a}$, when $x = x_{max} = L_0$ and the solution singularity can be realized only on positive semi axis x>0. For the first time solution (3.1) in the cases n=2 and n=3 is obtained in [6].

2. When introducing into the right side of equation (1.5) the homogeneous friction force $-\mu u_i$ with coefficient $\mu > 0$ the exact solution for such a way modified equation (1.5) is obtained from (3.1) if to substitute in (3.1) the time variable by new variable $t \to \tau(t) = (1 - e^{-\mu t})/\mu$. Besides, it is obvious that if condition

$$\mu > \mu_{th} = \frac{1}{t_0} \qquad (3.2)$$

is satisfy, the solution (3.1) remains smooth for any arbitrarily large times. Actually, under this condition even in the limit $t \to \infty$, determinant $\det \hat{A}$ does not vanish, remaining positive, since the condition $\tau(t) < t_0$ is preserved in this limit. For example, for the considered above evaluation at n=1 condition (3.2) has the following form $\mu > \mu_{th} = \frac{a}{L\sqrt{e}}$. At the same time the noted modification of solution (3.1) exactly satisfies vortex 3D equation (1.6) when introducing member $-\mu \omega_i$ into the right side of (1.6).

For any numerical solution of the Navier-Stokes equation, which contains a viscosity force of the form $\nu \Delta \vec{u}$ in the right side, the truncation on some large wave numbers $k = k_{max}$ is inevitably introduced. This leads to the appearance of an effective homogeneous friction with coefficient $\mu = k_{max}^2 \nu$ and if the condition (3.2) is violated, the occurrence of a numerical instability leading to the limitation of the predictability limits of the corresponding calculations is a fortiori possible.

In [27] it is noted that even in the numerical integration of the Euler equation for an ideal non viscous medium, some effective friction still appears, which, apparently, can also be simulated by the approximation of homogeneous friction.

3. If we replace $u_i \to u_i + V_i(t)$ in (1.5), then solution (3.1) will still exactly satisfy equations (1.5) and (1.6) if we replace

$$\vec{x} \to \vec{x} - \vec{B}(t), \vec{B}(t) = \int_0^t dt_1 \vec{V}(t_1)$$

in (3.1).

Let us now assume that velocity field $\vec{V}(t)$ is a random Gaussian field of white noise type, satisfying the conditions (angular brackets hereinafter mean statistical averaging):
$\langle V_i(t) \rangle = 0; \langle V_i(t+\tau) V_j(t) \rangle = 2\nu \delta_{ij} \delta(\tau)$. After statistical averaging of this modification of expression (3.1) we obtain for the average velocity an expression that remains smooth for any time intervals and has the following form:

$$\langle u_i \rangle = \int d^n \xi u_{0i}(\vec{\xi}) |\det \hat{A}| \frac{1}{(\sqrt{2\pi \langle \vec{B}^2(t) \rangle})^n} \exp\left[-\frac{(\vec{x} - \vec{\xi} - t\vec{u}_0(\vec{\xi}))^2}{2\langle \vec{B}^2(t) \rangle}\right] \quad (3.3)$$

$$\langle \vec{B}^2(t) \rangle = 2nt\nu$$

If we consider the Gaussian velocity $\vec{V}(t)$, but with a finite correlation in time (another then delta-function used to obtain (3.3)) the value $\langle \vec{B}^2(t) \rangle$ in (3.3) must be correspondingly changed in (3.3) on some another function of time.

Thus, a smooth for any times solution of the Navier-Stokes equation for a viscous compressible medium in the form (3.3) is obtained. It takes place under condition of taking into account the relations (2.6) in (1.1) and (1.3), and it is permissible to use instead of term $\frac{\eta}{\rho} \frac{\partial^2 u_i}{\partial x_i^2}$ in the right side of (1.1) for simulating the viscosity force by a random Gaussian velocity field of white noise type. The later replacement gives the possibility of the decoupling for velocity and density fields in (1.5) and (1.2) in the limit of small viscosity $\eta \to 0$ when for a far out-of-equilibrium systems the second viscosity is relatively large $\eta \ll \varsigma$ and presentation of pressure in (2.6) become exactly the same as in formulas for pressure (81.4) in [7] and (1.95) in [8].

In limit $\nu \gg \frac{\eta}{\min \rho}$ this substitution seems justified, where value $\nu$ corresponds to the level of white noise simulating the effect of viscosity forces. This method of simulating the viscosity does not lead to the appearance of hyper viscosity effects (when instead of the Laplacian in the Navier-Stokes equation introduced is a viscosity force proportional to the Laplacian in a degree higher than one [1]).

4. Equation of continuity (1.2) has a structure which exactly coincides with equation for vortex field (1.6) in two-dimensional case, when in the right side of (1.6) the first term is absent. For solution (3.1) the vortex field in two-dimensional case exactly satisfies equation (1.6) and has the following form [9]:

$$\omega(\vec{x},t) = \int d^2 \xi \omega_0(\vec{\xi}) \delta(\vec{\xi} - \vec{x} + t\vec{u}_0(\vec{\xi})) \quad (3.4)$$

where $\omega_0(\vec{x})$ is an initial distribution of the vortex field on plane. In case of susbstitution $\omega_0(\vec{x}) \to \rho_0(\vec{x})$ in expression (3.4) gives the distribution for the density field (not only in two-dimensional, but also in three-dimensional case, if the integration is considered in three-dimensional space in (3.4) and all the vectors appearing in (3.4) are also considered as three-dimensional).

The exact solution of the three-dimensional Euler vortex equation (1.6) corresponding to the velocity field (3.1) has the following form (also see validation of this in Appendix B in [9]):

$$\omega_i(\vec{x},t) = \int d^3 \xi (\omega_{0i}(\vec{\xi}) + t\omega_{0j} \frac{\partial u_{0i}(\vec{\xi})}{\partial \xi_j}) \delta(\vec{\xi} - \vec{x} + t\vec{u}_0(\vec{\xi})) \quad (3.5)$$

where $\vec{\omega}_0 = rot\vec{u}_0$. The expression for enstrophy corresponding to solution (3.5) has the following form:

$$\Omega_3 \equiv \int d^3x \omega_i^2(\vec{x},t) = \int d^3\zeta (\omega_{0i} + t\omega_{0j}\frac{\partial u_{0i}}{\partial \zeta_j})^2 / \det \hat{A} \qquad (3.6)$$

The expressions for any higher single- and multi-point moments of the vortex field may be obtained similarly. For simplicity's sake, let us give them only for two-dimensional case, when, for example, single-point moments of an arbitrary order have the following form:

$$\Omega_{2(m)} = \int d^2x \omega^m = \int d^2\xi \frac{\omega_0^m(\vec{\xi})}{\det^{m-1} \hat{A}}; \Omega_{2(2m)} = \int d^2x \omega^{2m} = \int d^2\xi \frac{\omega_0^{2m}(\vec{\xi})}{\det^{2m-1} \hat{A}}; m = 1,2,3,... \qquad (3.7)$$

Thus, the exact solution of the closure problem, the main problem in the theory of turbulence (see [16]), is obtained for the case of a compressible medium. In particular, from (3.7) it follows that in limit $t \to t_0$ we have inequality $\Omega_{2(m)}^2 \cong O(\frac{1}{(t_0-t)^{2(m-1)}}) << \Omega_{2(2m)} \cong O(\frac{1}{(t_0-t)^{2m-1}})$, which is typical for realization of the high intermittency of turbulence [1].

Similarly, for the asymmetry and excess values we obtain the evaluations:

$s = b_3 / b_2^{3/2} \cong O(\frac{1}{(t-t)^2})$, $\frac{b_4}{b_2^2} - 3 \cong O(\frac{1}{(t_0-t)^2})$ since $\det \hat{A} \cong O(t_0-t)$ in the limit $t \to t_0$. At the same time, in accordance with [16], $b_2; b_3; b_4$ are central moments of the second, third, and fourth order for the vortex field (3.4).

Basing on solution (3.4) we may obtain the following evaluation of the finite energy dissipation rate in two-dimensional compressible turbulent flow within the limit of vanishingly low viscosity, when $t \to t_0$:

$$\varepsilon_d = \lim_{\nu \to 0} \nu \|\langle \omega \rangle^2\| = \frac{1}{8t_0\pi} \int d^2\xi \Omega_0(\vec{\xi}) \qquad (3.8)$$

$$\|\omega_0(\vec{\xi})\omega_0(\vec{\xi}_1)\| = \delta(\vec{\xi}_1 - \vec{\xi})\Omega_0(\vec{\xi}) \qquad (3.9)$$

In (3.8), (3.9) the double bracket operator denotes the statistical averaging over the ensemble of initial fields realization and the limit in (3.8) is obtained under condition of zero correlation of initial velocity and vortex field $\|\omega_0 u_{0i}\| = 0$, which follows from assumption about statistical homogeneity of initial fields.

The intermittency of turbulence also manifests itself in the observed deviation of the scaling law for the structure function of velocity field $\|(\delta u(l))^p\| \cong O(l^{\zeta_p})$ from exponent $\zeta_p = p/3$ obtained in the Kolmogorov-Obukhov theory [1]. For example, from exact solution (3.1) in one-dimensional (for simplicity's sake) case we obtain a different from $\zeta_2 = 2/3$ value $\zeta_2 = 1$ as within limit $t << t_0$ we have the following from (3.1):

$$\int_{-\infty}^{\infty} dx (\delta u)^2 \cong 2tl \int_{-\infty}^{\infty} dx u_0 (\frac{du_0}{dx})^2 + O(l^2), \qquad (3.10)$$

Following from (3.10) evaluation $\zeta_2 = 1$, actually, coincides well with the values of indices $\zeta_2^{11} = 0.952 \pm 0.004; \zeta_2^{\perp} = 0.977 \pm 0.008$, which are obtained from a numerical simulation of supersonic isothermal turbulence in [28] on the basis of the 3-D Euler equation for a compressible medium.

### 4. Energy spectra of compressible turbulence

On the base of solution (3.3) to the Navier-Stokes equation (1.1)-(1.3) it is possible to obtain representation for all multipoint moments of velocity, density and pressure fields, which gives the solution of the main problem of the turbulence theory [16] (see chapter 3.3 on the pages 176-177 in [16]). Indeed, in [16] it is noted that for the case of a compressible medium, "…this common problem is too difficult, and the approach to its full solution is not yet visible" [16]. Here we consider the example in the compressible case of the solution for two-point moments of velocity field and corresponding energy spectrum which are of the most importance in turbulence theory.

The correlation tensor for the n-dimensional (n=1, 2, 3) velocity field is defined as:

$$R_{ij}(\vec{l}) = \frac{1}{L^n}\int d^n x \tilde{u}_i(\vec{x}+\vec{l},t)\tilde{u}_j(\vec{x},t); \tilde{u}_i = \langle u_i(\vec{x},t)\rangle \quad (3.11)$$

$$L^n = (\frac{1}{2\sqrt{\pi}})^n P_0^2/E_0; P_0 = \int d^n x |\vec{u}_0(\vec{x})|; E_0 = \int d^n x \tilde{u}_0^2(\vec{x})$$

Let us consider (3.11) for the solution of Navier-Stokes equation in the form (3.3) for the steady limit $t\mu \gg 1$, when before in (3.3) we replace $t \to \frac{1-\exp(-t\mu)}{\mu}$ after introduction homogeneous friction with coefficient $\mu > 0$. The correlation tensor (3.11) in this case of presentation for (3.3) is reduced to the formula

$$R_{ij}(\vec{l}) = \frac{1}{(4\pi L^2\langle\vec{B}^2\rangle)^{n/2}}\int d^n\xi\int d^n\xi_1 u_{0i}(\vec{\xi})u_{0j}(\vec{\xi}_1)\det\hat{A}(\vec{\xi})\det\hat{A}(\vec{\xi}_1)\exp(-\frac{(\vec{l}+\vec{a}(\vec{\xi},\vec{\xi}_1))^2}{4\langle\vec{B}^2\rangle})\quad(3.12)$$

In (3.12) vector $\vec{a}(\vec{\xi},\vec{\xi}_1) = -\vec{a}(\vec{\xi}_1,\vec{\xi}) = \vec{\xi}_1 - \vec{\xi} + \frac{1}{\mu}(\vec{u}_0(\vec{\xi}_1) - \vec{u}_0(\vec{\xi}))$ and in accordance with (3.3) we also have $\langle\vec{B}^2\rangle = 2n\nu/\mu$. From (3.12) the symmetry $R_{ij}^u(\vec{l}) = R_{ji}^u(-\vec{l})$ is stated.

The spectrum energy tensor corresponding to (3.11) is defined as

$$F_{ij}(\vec{k}) = \frac{1}{(2\pi)^n}\int d^n l R_{ij}(\vec{l})\exp(i\vec{k}\vec{l}) \quad (3.13)$$

Indeed, the energy (on the mass unit) is defined as:

$$e = \frac{1}{2}\int d^n k F_{ii}(\vec{k}) \equiv \int_0^\infty dk E(k) \quad (3.14)$$

For example, in the isotropic 2D case there is relation $F_{ii}(k)\pi k = E(k)$ and for 3D case also takes place $F_{ii}(k)4\pi k^2 = E(k)$.

From (3.13) and (3.12) it is possible to obtain representation

$$F_{ij}(\vec{k}) = F_{ji}(-\vec{k}) = \frac{1}{L^n}I_i(\vec{k})I_j(-\vec{k})\exp(-k^2\langle\vec{B}^2\rangle);$$

$$I_i(\vec{k}) = \int d^n x u_{0i}(\vec{x})\exp(i\vec{k}(\vec{x}+\frac{1}{\mu}\vec{u}_0(\vec{x})))\det\hat{A}(\vec{x}) \quad (3.15)$$

For simplicity, first of all let us consider (3.15) with n=2 for the 2D initial velocity field (in polar coordinate $(r,\varphi)$ system):

$$u_{0r} = 0; u_{0\varphi}(r) = a\exp(-\frac{r^2}{2L_0^2}) \qquad (3.16)$$

For the velocity (3.16) in (3.15) $L = L_0$ and in (3.2) $\mu_{th} = a/L_0$ (by using (3.7) from [9] in the case with n=2). In this case from (3.15), (3.16) it is possible to obtain:

$$F_{\varphi\varphi}(k) = \frac{I_\varphi(k)I_\varphi(-k)}{L_0^2}\exp(-4\nu k^2/\mu); \qquad (3.17)$$

$$I_\varphi(k) = a\int_0^{2\pi}d\varphi\int_0^\infty dr\, r\exp(-\frac{r^2}{2L_0^2})(1-\frac{\mu_{th}^2}{\mu^2}\exp(-\frac{r^2}{L_0^2}))\exp(ikS(r,\varphi));$$

$$S(r,\varphi) = r\cos\varphi - \frac{\mu_{th}}{\mu}L_0\exp(-\frac{r^2}{2L_0^2}) \qquad (3.18)$$

For the estimation of integral (3.18) in the limit of large wave number $k$ we use the method of stationary phase when

the conditions $\frac{\partial S}{\partial r} = \frac{\partial S}{\partial \varphi} = 0$ are take place for

$$r = r_0 = L_0\sqrt{2}\ln^{1/2}(\mu_{th}/\mu); \varphi = \varphi_0, \operatorname{ctg}\varphi_0 = -\frac{\mu_{th}r_0}{\mu L_0}\exp(-\frac{r_0^2}{2L_0^2}) \qquad (3.19)$$

As an example, for the case $\mu < \mu_{th}$ and in the limit $1 - \frac{\mu_{th}}{\mu} \ll 1$ we obtain here the estimation of (3.18) for the inertial range of wave numbers (when $k^2 4\nu/\mu \ll 1$ in (3.17))

$$1/L_0 \ll k \ll \sqrt{\operatorname{Re}}/L_0; \operatorname{Re} = L_0 U_0/\nu, U_0 = L_0\mu/4 \qquad (3.20)$$

For the wave numbers range (3.20) from (3.17) the 2D energy spectrum is obtained:

$$E(k) = \frac{C_E}{k^3}; C_E = \frac{\pi^3 a^2}{256 L_0^2}\ln^{-3}(\mu_{th}/\mu) \qquad (3.21)$$

This representation of energy spectrum (3.21) is obtained only on the base of exact solution (3.3) of the compressible NS equation (1.1), (1.3) and has the same scaling exponent as usually considered in 2D turbulence theory [29] when also $E(k) \cong O(1/k^3)$ in the inertial range of wave numbers is obtained, but only for incompressible case and not on the base of the solution of Navier-Stokes directly.

For 1D case with n=1 in (3.15) it is also possible to obtain simple example when initial velocity field in (3.15) takes form $u_0(x) = a\exp(-x^2/2L_0^2)$, when $L = L_0$ and $\mu_{th} = a/L_0\sqrt{e}$. In the case $\mu > \mu_{th}$ from (3.15) (by using of the method of integrating by parts) the energy spectrum takes the form:

$$E(k) = \frac{C_1(k)}{k^4}\exp(-\frac{2\nu k^2}{\mu});$$

$$C_1(k) = \frac{a^2(1+\mu_{th}^2/\mu^2-(1-\mu_{th}^2/\mu^2)\cos 2kL_0)}{\pi e L_0^3(1-\mu_{th}^2/\mu^2)^2} \qquad (3.22)$$

In the limits $\mu \gg \mu_{th}$ and $kL_0 \ll 1$; $\mathrm{Re} = U_0 L_0/\nu \gg 1$, $U_0 = L_0\mu/2$ from (3.22) we have scaling law for one-dimensional energy spectrum $E(k) = 4a^2/\pi e L_0 k^2$ which is known for the 1D Burgers turbulence[30] (see Fig.27 in [30] for the spectrum of velocity potential in the inertial range, where it is proportional to $O(1/k^4)$ which is corresponding to $E(k) \cong O(1/k^2)$).

In the inertial range of wave numbers (for (3.20), where $U_0 = L_0\mu/2$ in the case n=1) in the case $\mu < \mu_{th}$ we have in the limit $1 - \mu/\mu_{th} \ll 1$:

$$E(k) = C_2/k^3;$$
$$C_2 = \mu^3 L_0 / 2a\sqrt{1 - \mu/\mu_{th}} \quad (3.23)$$

Thus, in 1D case in (3.23) we have the same scaling of energy spectrum $E(k) \cong O(1/k^3)$ as in (3.21) for 2D case where also the condition $\mu < \mu_{th}$ is used.

Also in the inertial range of wave numbers on the base of stationary phase method (as we obtain (3.23) from (3.15) for n=1) in the case of $\mu = \mu_{th}$ the 1D energy spectrum is obtained in the form:

$$E(k) = C_3/k^{8/3};$$
$$C_3 = \frac{2a^2 I_{1/3}^2}{\pi e 3^{1/3} L_0^{5/3}}, \quad I_{1/3} = \int_0^\infty dx \frac{\exp(-x^2)}{x^{1/3}} \quad (3.24)$$

It is interesting to note that the same type spectrum as in (3.24) (for its frequency representation when $k = \omega/\overline{V}$, $\omega$ - frequency and $\overline{V}$ the averaged one dimensional velocity) was obtained in experimental date [31-33]. Indeed, in [31, 32] it was obtained measurements of the spectrum of the radio waves phase, which have the same as in (3.24) scaling law $O(1/\omega^{8/3})$. From the other side the spectrum of phase has the same scaling as for the spectrum of refractive index fluctuations, when the last is completely determined by one-dimensional fluctuations of velocity in the case with zero temperature fluctuations (see (26.69) and (26.76) in [33]).

## Conclusion

In this paper the new analytic solution to the Navier-Stokes is obtained. It is shown that condition (3.2) makes it possible to establish, the previously not taken into account, new relation between the procedure of truncation at some maximal wave number $k_{max}$ (or corresponding to $k_{max}$ minimal size of the numerical grid $l_{min} \approx 1/k_{max}$), which is unavoidable in any numerical experiment, and the initial conditions. The latter determine the value of the singularity time $t_0$, at which the predictability limit can be qualitatively and quantitatively increased by eliminating the singularity of the solution, only when the condition (3.2) is take place. This means that the resulting positive solution of the Millennium prize problem (see [13]) generalization for the case of compressible divergent flows provokes not only purely mathematical interest, as it has been presented earlier (see [1]), but can help to improve the computational codes used, for example, in the formation of short-term forecasts of adverse atmospheric phenomena.

It is shown that, near the singularity of the solution, the exact representations for the values of the second, third, and fourth moments of the vortex field indicate a significant deviation from that in the case of Gaussian statistic. In this case, as opposed to the evaluations obtained earlier (see, for example, [1]), just on the basis of the above exact solution established is the finite critical value of the energy dissipation velocity in the limit of vanishingly small viscosity, and also obtained is the evaluation of the structure function scaling index which is well consistent with a numerical experiment [28].

On the basis of the exact solution for velocity (3.1) and vortex (3.4), (3.5) fields, any single-point and multi-point moments of these fields can be analytically obtained, what gives the closure problem exact solution (see common energy spectra (3.15) and some examples of in in n=2 and n=1). To solve this fundamental problem of the turbulence theory, various interesting physical and mathematical approximate methods and approaches have been proposed earlier [34], the applicability analysis of which can now be carried out on the basis of the exact solution obtained above for the case of a compressible medium.

We thank E.A. Kuznetsov, A.G. Kritsuk, P. Lushnikov, I. Procaccia and
G. Falkovich for useful discussions.

The study is supported by RSF, project No. 14-17-00806П